# Identification of DNA Bases Using Nanopores Created in Finite-Size Nanoribbons from Graphene, Phosphorene, and Silicene

Matthew B. Henry, Mukesh Tumbapo, and Benjamin O. Tayo

*Department of Engineering and Physics, University of Central Oklahoma, Edmond, OK 73034, USA*

Author to whom correspondence should be addressed: btayo@uco.edu

**ABSTRACT**

Graphene's success for nanopore DNA sequencing has shown that it is possible to explore other potential single- and few-atom thick layers of elemental 2D materials beyond graphene (e.g., phosphorene and silicene), and also that these materials can exhibit fascinating and technologically useful properties for DNA base detection that are superior to those of graphene. Using density functional theory (DFT), we studied the interaction of DNA bases with nanopores created in finite-size nanoribbons from graphene, phosphorene, and silicene. Due to the small size of DNA bases, the bases interact with only a small section of the nanoribbon, hence using a finite-size model is appropriate for capturing the interaction of bases and 2D membrane materials. Furthermore, by using a finite-size model, our system is approximated as a molecular system, which does not require a periodic DFT calculation. We observe that binding energies of DNA bases using nanopores from phosphorene and silicene are similar, and generally smaller compared to graphene. This shows that minimal sticking of DNA bases to pore is expected for phosphorene and silicene devices. Furthermore, nanopores from phosphorene and silicene show a characteristic change in the density of states for each base. The band gaps of phosphorene and silicene are significantly altered due to interaction with DNA bases compared to graphene. Our findings show that phosphorene and silicene are promising alternatives to graphene for DNA base detection using advanced detection principles such as transverse tunneling current measurement.

## I. INTRODUCTION

Research into methods of deoxyribonucleic acid (DNA) sequencing has been improving with the goal of providing fast and cheap sequencing of longer strands as accurately as possible. One possible way of improving the process is utilizing solid-state devices made from two-dimensional (2D) materials. Graphene remains the most widely studied 2D material for DNA sequencing. Graphene nanopores and nanogaps have been successfully used for DNA sequencing.[1,2] While the single-layer nature of graphene provides the optimal thickness (0.34 nm) for single-base resolution,[3] the major hindrance is the hydrophobic nature of graphene's surface. Because of the strong π-π interactions between graphene and the DNA, bases stick to its surface[4,5] leading to a significant reduction in translocation speed due to pore clogging.[6] Furthermore, the coexistence of different bases on the surface and pore makes single-base discrimination difficult.[6] Another issue is the problem of orientational fluctuations of nucleobases during DNA translocation through a graphene nanopore. This can give rise to overlapping current contributions from different bases.[7] It has been shown that the ionic blockade signal shows noise for DNA translocation through a single-layer graphene nanopore.[5] The origin of this noise has been attributed to the atomic thickness of the pore. It is notable that a nanopore in a three-layer graphite structure, which has a thickness ~1 nm, shows a better signal-to-noise ratio (SNR).[5,8] The lack of a band gap in pristine graphene makes it undesirable for use in electronic-based detection modalities such as tunneling current or field-effect transistors (FETs).[9]

Despite the challenges of graphene, graphene's success for nanopore DNA sequencing has shown that it is possible to explore other potential single- and few-atom thick layers of 2D materials beyond graphene, and also that these materials can exhibit fascinating and technologically useful properties for DNA base detection that are superior to those of graphene. One such material is molybdenum disulfide ($MoS_2$). Among the large family of 2D transition metal dichalcogenides (TMDs), $MoS_2$ is the most widely studied for sequencing applications, mainly due to the ease of fabrication of $MoS_2$ devices.[10-12] Several theoretical and experimental studies have demonstrated sequencing using single-layer nanopores and nanoribbons from $MoS_2$.[9,10,13-15] These studies reveal that $MoS_2$ performs better than graphene. For instance, improved SNR, non-stickiness of DNA to $MoS_2$ surface, and the presence of an intrinsic band gap makes it suitable for use in advanced sequencing devices such as FETs or the nanochannel device.[9,10] Another 2D material that has been explored for DNA sequencing is hexagonal boron nitride (hBN). hBN is less hydrophobic than graphene which can minimize the hydrophobic interaction that impedes the DNA translocation through the constructed nanopores. Moreover, the thickness of hBN is comparable to the spacing between nucleotides (0.32 – 0.52 nm) in single-stranded DNA (ssDNA).[16] It also shows other advantages over graphene in terms of its insulating property in high-ionic-strength solution and fewer defects made during the manufacturing process.[17] Experimental and theoretical studies show that hBN can reach spatial resolutions for DNA sequencing close to those of graphene.[18-19] The fundamental properties of DNA nucleobases and hBN sheets remain unchanged upon adsorption, which suggests its promising application for DNA research.[20] The above consideration shows that hBN is a nice alternative to graphene and could be used in combination with graphene for advanced functional devices for DNA sequencing with higher SNR.[19]

While tremendous success has been reported with the use of 2D materials (e.g., graphene, hBN, $MoS_2$, and $WS_2$) for DNA sequence, there are still fundamental limitations related to the high translocation speed during DNA sequencing that has to be resolved.[9,10,21] At this stage, it is not clear what solid-state nanopore material will be able to meet the challenges for single-base resolution. Thus, it is crucial to continue to carry out explorative studies to identify new nanopore materials that could potentially emerge as the best candidate materials. Two such candidate materials are phosphorene[3,22] and silicene.[11,23] The puckered honeycomb structure of phosphorene and buckled honeycomb structure of silicene makes these materials thicker than graphene, hence superior materials for single-base DNA sequencing in terms of signal-to-noise ratio intensity while maintaining monolayer properties.[9] Furthermore, phosphorene and silicene share many of the remarkable properties of graphene including high carrier mobility and tunable optical properties.[23-26] Both phosphorene and silicene have direct band gaps that are dependent on thickness, making them ideal for advanced base detection principles that probe modulations in electronic and optoelectronic properties.[23,27] Moreover, the hydrophilicity and biocompatibility of phosphorene makes it ideal for applications in biosensing.[28,29]

DNA sequencing using nanoribbons from phosphorene was recently demonstrated using periodic density functional theory (DFT) calculations.[29,30] These studies were based on probing modulations in electronic properties due to translocation of DNA bases through a nanogap in phosphorene[29] or modulations caused by physisorption of bases onto the surface of phosphorene nanoribbons.[30] In this article, using DFT calculations, we studied the interaction of DNA bases with nanopores created in finite-size nanoribbons from graphene, phosphorene, and silicene. Due to the small size of DNA bases (~0.6 – 0.8 nm), the bases interact with only a small section of the nanoribbon, hence using a finite-size model is appropriate for capturing the interaction between bases and 2D membrane materials. Furthermore, by using a finite-size model, our system is modeled as a molecular system, which does not require a periodic DFT calculation. We observe that binding energies of DNA bases using nanopores from phosphorene and silicene are similar, and generally smaller compared to graphene. This shows that minimal sticking of DNA bases to pore is expected for phosphorene and silicene devices compared to graphene.[4,5] Furthermore, nanopores from phosphorene and silicene show a characteristic change in the density of states for each base. The band gap of phosphorene and silicene are significantly altered due to interaction with DNA bases compared to graphene. Our findings show that phosphorene and silicene are promising alternatives to graphene for DNA base detection using advanced detection principles such as transverse tunneling current measurement.[31]

## II. MATERIALS AND COMPUTATIONAL METHODS

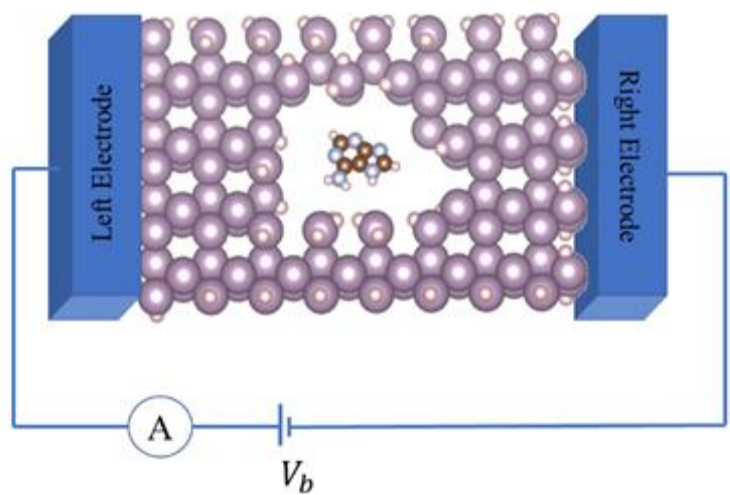

**FIG. 1.** Schematic of nanopore sensing device. Translocation of DNA base through pore causes variations in in-plane current. The in-plane current flows from left to right, that is perpendicular to the DNA backbone during DNA traversal.

Our model is based on detecting variations in the in-plane current[7,31] through a nanopore created in a finite-size nanoribbon membrane due to traversal of a DNA base as shown in **Figure 1**. First, we performed DFT studies for the individual DNA bases to determine the maximum width of each base. This was determined to be 0.754 nm for guanine (G), 0.648 nm for adenine (A), 0.560 nm for cytosine (C), and 0.590 nm for thymine (T). Then nanopores (1 – 1.2 nm) were created in each of the three membrane materials, that is graphene, phosphorene, and silicene. Our pore sizes are in the 1 - 2 nm range used in previous studies.[7,31] Such a pore size is good enough to allow nucleotides (size ∼ 1.0 to 1.2 nm) to pass through.[29] However, for the detection of double-stranded DNA (dsDNA), a pore diameter of 2.3 nm has been shown to be the smallest and most efficient pore size.[4,9] Nucleotides serve as monomeric units of DNA and consist of a base, a sugar, and a phosphate group. Since the sugar and phosphate groups are the same for each nucleotide, we shall consider only the interaction of the four target bases (G, A, C, and T) with nanopores from graphene, silicene, and phosphorene in this study. This approximation has been used in previous computational studies.[7,9]

We will refer to our finite-size armchair nanoribbon systems using the following abbreviations: GNP (graphene nanopore), PNP (phosphorene nanopore) and SNP (silicene nanopore). Due to the small size of DNA bases, the bases interact with only a small section of the nanoribbon, hence using a finite-size model is appropriate for capturing the interaction of bases and 2D membrane materials. Furthermore, by using a finite-size model, our system is approximated as a molecular system, which does not require a periodic DFT calculation.

For each membrane material, nanopores were created by deleting central atoms. Hydrogen passivation[9,30] was used to terminate the dangling bonds at the edges and pore. Structural minimization was performed for the individual bases and membranes separately, and then for the combined system of nanopore and bases. All structural relaxation calculations were performed at the B3LYP level of theory using the 6-31G (d, p) basis set, with a force convergence cutoff of 0.02 eV/Å.[30] The combination of the B3LYP hybrid-GGA (generalized gradient approximation) functional with the 6-31G (d, p) basis set provides a reasonable basis set size for performing geometry optimization calculations. All calculations were performed using the GAUSSIAN 16 software package.[32] Computational resources were provided by the University of Central Oklahoma Buddy Supercomputing Center.[33]

To evaluate the potential of phosphorene and silicene for DNA base detection, we computed the electronic structure changes induced by the presence of DNA bases inside the nanopore using DFT. Three evaluation metrics, namely, energy band gap, binding energy, and electronic density of states (DOS) were considered. As graphene is the most widely studied 2D material for DNA sequencing, our results for phosphorene and silicene were benchmarked against graphene. The energy band gap is given by

$$E_{gap} = E_{HOMO} - E_{LUMO} \quad (1)$$

where $E_{HOMO}$ is the energy of the Highest Occupied Molecular Orbital (HOMO) state, and $E_{LUMO}$ the energy of the Lowest Unoccupied Molecular Orbital (LUMO) state. The binding energy for the different systems was calculated as follows

$$E_{bind} = E_{nanopore+base} - E_{nanopore} - E_{base} \quad (2)$$

Here, $E_{bind}$ is the binding energy for the system, $E_{nanopore+base}$ is the total energy of combined system (nanopore and base), $E_{nanopore}$ is the energy of the nanopore, and $E_{base}$ the energy of a single DNA base. In a real device, the tunneling current signal is computed numerically from the integrated DOS.[7,34]

$$I(E, V_b) = \frac{e}{\pi \hbar} \int_{-\infty}^{E} DOS(E - E') dE' \quad (3)$$

This suggest that a modulation in the DOS of the pristine nanopore system should produce a modulation in the tunneling current. In this study, instead of performing the full calculation of the tunneling current which requires the non-equilibrium Green's Function formalism,[7,29,30,34] we shall focus on studying variations in the DOS of the pristine nanopore due to interaction with DNA bases.[9] The DOS was computed using the Lorentzian function as[35]

$$DOS(E) = \sum_{i=0}^{N} \frac{\Gamma}{\Gamma^2 + (E - E_i)^2} \quad (4)$$

where $E$ is the energy, $E_i$ are the molecular energy levels obtained from the DFT calculation, and $\Gamma$ is the linewidth of the Lorentzian spectral function. In our calculations, we used $\Gamma$ = 25 meV as this corresponds to the linewidth at room temperature.[35]

## III. RESULTS

**III.1 Graphene Nanopore (GNP).** Our finite-size GNP system has 15 dimer lines along the zigzag direction and 4 translational periods along the armchair direction. The optimized length along the zigzag direction is 1.91 nm, and the optimized length along the armchair direction is 1.80 nm. The pore diameter is approximately 1.07 nm. **Table I** shows the band gap and binding energies for each system.

**TABLE I.** Energy gap and binding energy for GNP systems.

| System | $E_{gap}$ (eV) | $E_{bind}$ (eV) |
|---|---|---|
| **GNP + G** | 0.228 | -0.888 |
| **GNP + A** | 0.230 | -0.936 |
| **GNP + C** | 0.230 | -1.063 |
| **GNP + T** | 0.231 | -0.871 |

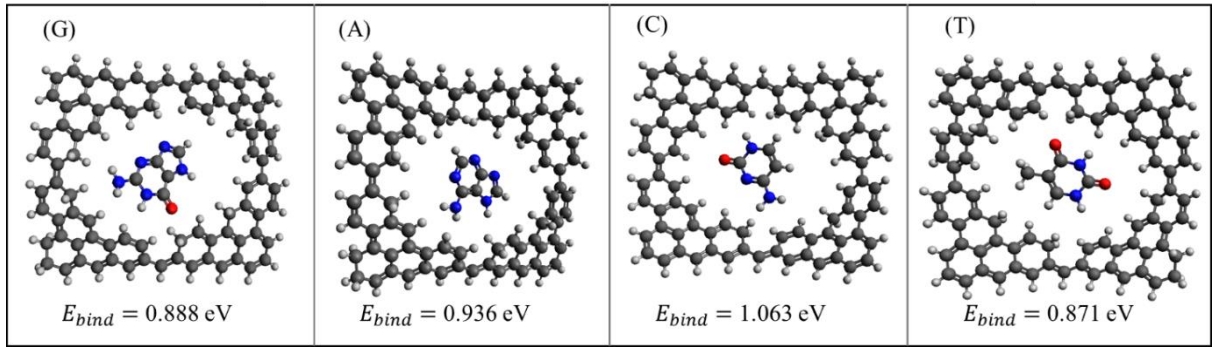

**FIG. 2.** Optimized structures of the GNP + nucleobase systems. The absolute value of the computed binding energies are shown for each base.

The band gap of pristine GNP is 0.221 eV. **Table I** shows that the band gap of GNP is not significantly affected by the presence of DNA bases. Each base slightly opens the band gap of the pristine GNP. Base T most significantly affects GNP and gives rise to an increase in band gap of 0.01 eV. The change in band gap is highest in T and lowest in G and follows the order T > C = A > G. The optimized geometry for each GNP system is shown in **Figure 2** along with the corresponding binding energies. Binding energies for our GNP calculation follow the order C > A > G ~ T. The variation of binding energy and energy gap for each base is shown in **Figure 3**.

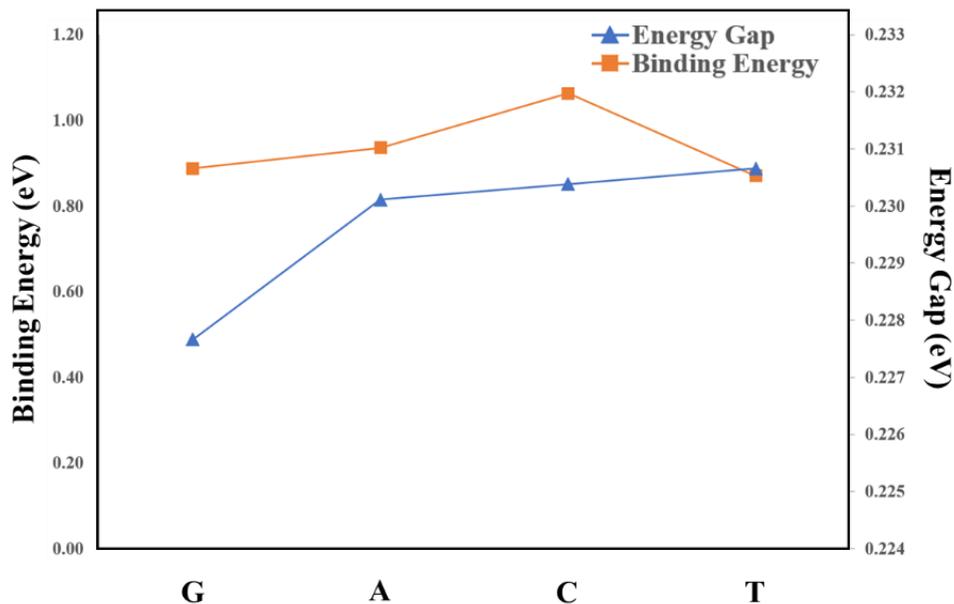

**FIG. 3**. Binding Energy (left axis) and Energy Gap (right axis) of the pristine GNP system with each DNA base placed in the pore.

**Figure 4** shows the DOS for pristine GNP and pristine GNP with DNA bases. The change is DOS is not very remarkable. From a visual inspection of the DOS plot, we observe that the change in DOS follows the sequence G > C > A > T.

The small variations in energy band gap and DOS of GNP due to interaction with DNA bases means that GNP is not suitable for detecting DNA bases using the in-plane transverse current method.[9,31] Also the binding energies of bases with GNP is relatively large. This indicates that the individual bases can easily stick to the surface of the graphene nanopore. Sticking to the surface can cause significant issues with base detection such as reduction in translocation speed due to pore clogging.[6]

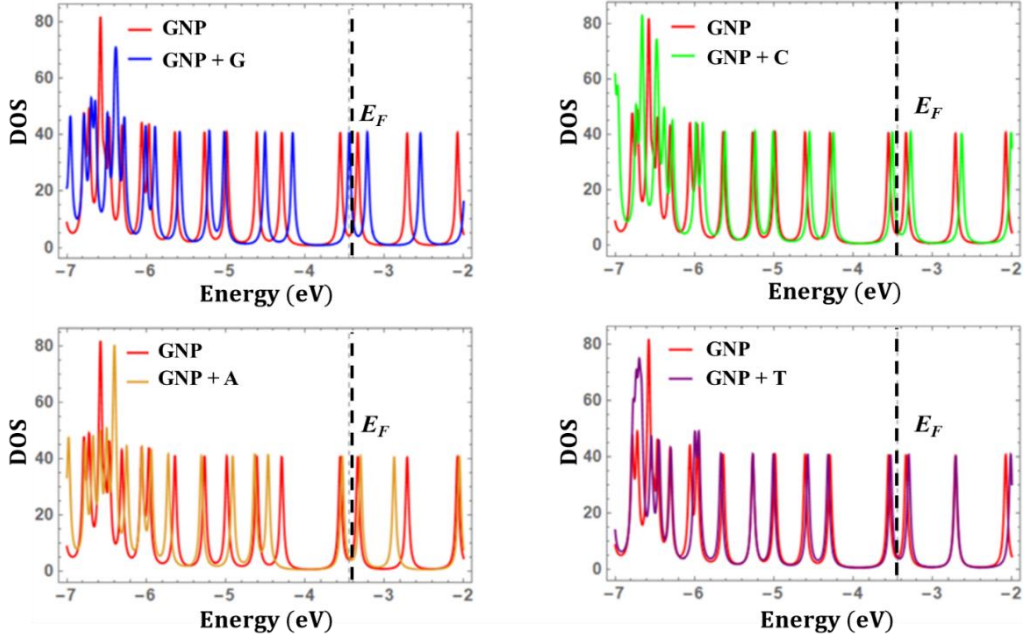

**FIG. 4**. Density of States of pristine GNP and pristine GNP + base systems. $E_F$ is the Fermi energy of pristine GNP.

**III.2 Phosphorene Nanopore (PNP).** Our finite-size PNP system has 16 dimer lines along the zigzag direction and 4 translational periods along the armchair direction. The optimized length along the zigzag direction is 2.48 nm, and the optimized length along the armchair direction is 1.20 nm. The pore diameter is approximately 1.03 nm. **Table II** shows the band gap and binding energies for each system.

Unlike the GNP, the band gap of each PNP system is significantly affected by the presence of bases. The pristine band gap for the PNP is 3.070 eV. We remark here that the band gap for the finite-size PNP system is much higher than the ~0.87 eV reported for an infinitely long phosphorene nanoribbon of identical width[30]. This difference is due to the finite size of our system, and the fact that the band gap of 0.87 eV was determined for a phosphorene nanoribbon with no pore.

From **Table II**, base A most significantly affects PNP and gives rise to a reduction in band gap by 0.281 eV compared to pristine PNP. Unlike other bases, base G slightly opens the band gap of PNP by 0.013 eV. The change in band gap is highest in A and lowest in G and follows the order A > T > C > G.

**TABLE II**. Energy gap and binding energy for PNP systems.

| System | $E_{gap}$ (eV) | $E_{bind}$ (eV) |
|---|---|---|
| **PNP + G** | 3.083 | -0.395 |
| **PNP + A** | 2.789 | -0.307 |
| **PNP + C** | 3.046 | -0.405 |
| **PNP + T** | 3.025 | -0.207 |

The optimized geometry for each PNP system is shown in **Figure 5** along with the corresponding binding energies for the different bases. The binding energy is a key parameter that describes the strength of the interaction between the DNA base and PNP. In GNP, the binding energies follow the sequence C > A > G > T. For PNP, the binding energies follow the sequence C > G > A > T. The variation of binding energy and energy gap for each base is shown in **Figure 6**.

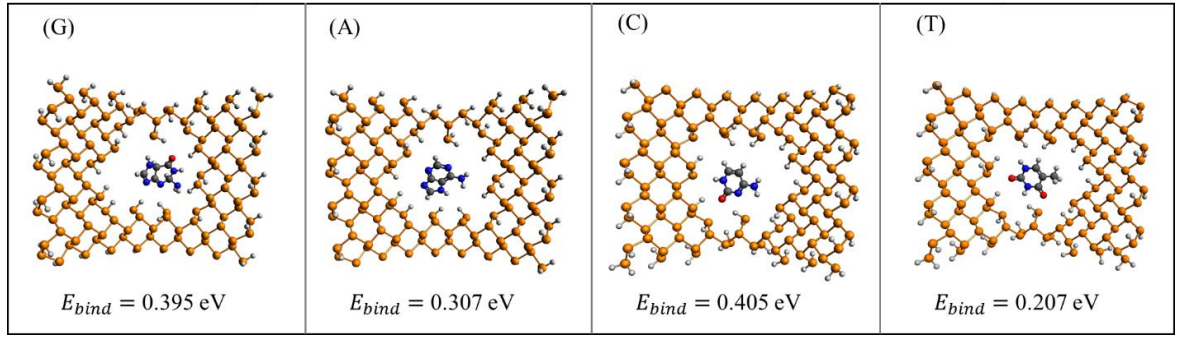

**FIG. 5**. Optimized structures of the PNP + nucleobase systems. The absolute value of the computed binding energies are shown for each base.

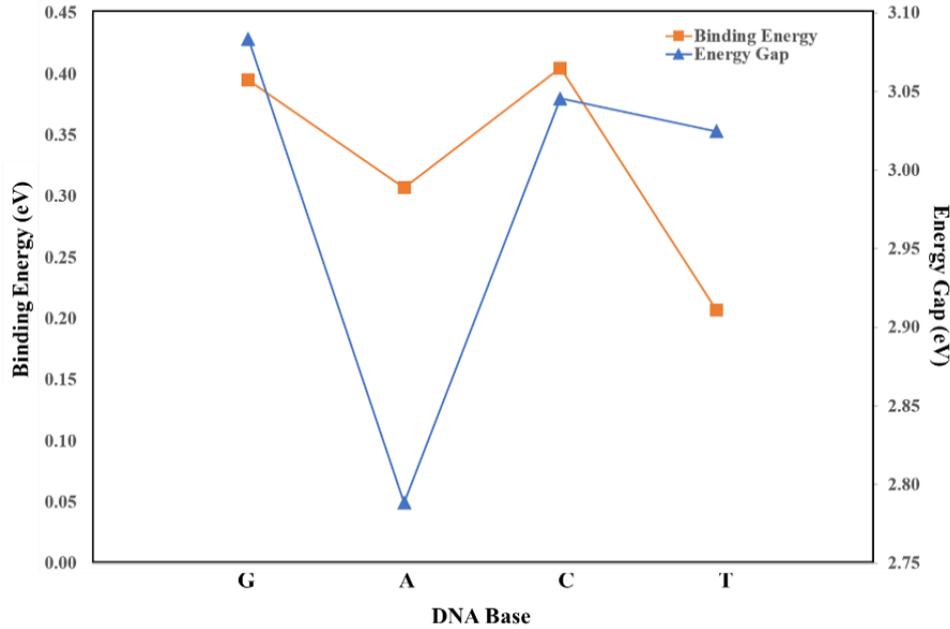

**FIG. 6**. Binding Energy (left axis) and Energy Gap (right axis) of the pristine PNP system with each DNA base placed in the pore.

**Figure 7** shows the DOS for pristine PNP and pristine PNP with DNA bases. The change is DOS is larger compared to GNP. From a visual inspection of the DOS plot, we observe that the change in DOS follows the same sequence as the change in band gap, that is A > T > C > G.

The large variations in band gap energy and DOS, and small binding energies of PNP due to interaction with DNA bases makes PNP a better material compared to GNP for DNA base detection using the in-plane transverse current method.[31]

**III.3 Silicene Nanopore (SNP).** Our finite-size SNP system has 13 dimer lines along the zigzag direction and 4 translational periods along the armchair direction. The optimized length along the zigzag direction is 2.57 nm, and the optimized length along the armchair direction is 2.58 nm. The pore has a diameter of approximately 1.20 nm. Due to the buckled structure of silicene, creating a perfectly circular pore was very challenging. Our calculated Si-Si bond length after a geometry optimization calculation was in the range 0.228 nm to 0.236 nm, which agrees nicely with the experimental value of 0.228 nm.[36,37] Furthermore, our calculated out of plane height for the buckled structure was found to be 0.045 nm, which also agrees nicely with the experimental value of 0.044 nm.[38]

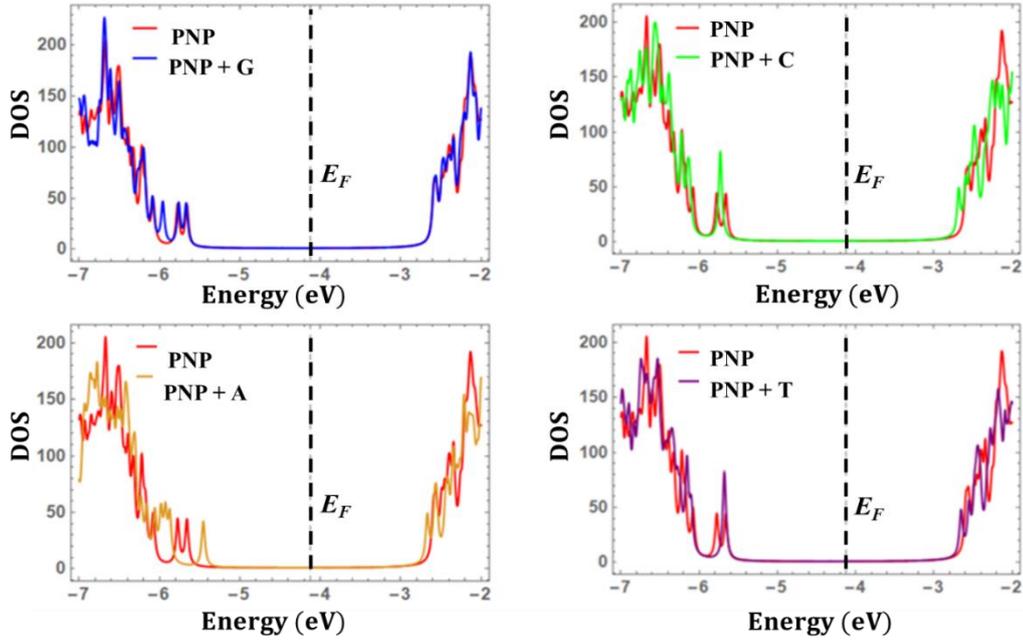

**FIG. 7**. Density of States of pristine PNP and pristine PNP + base systems. $E_F$ is the Fermi energy of pristine PNP.

**Table III** shows the band gap and binding energies for each system. The band gap of the pristine SNP system was calculated to be 2.219 eV. Base A most significantly affects SNP and gives rise to a reduction in band gap by 0.151 eV compared to pristine SNP. All bases give rise to a reduction in the band gap for SNP. The change in band gap is highest in A and lowest in T and follows the order A > C > G > T.

**TABLE III.** Energy gap and binding energy for SNP systems.

| System | $E_{gap}$ (eV) | $E_{bind}$ (eV) |
|---|---|---|
| **SNP + G** | 2.144 | -0.234 |
| **SNP + A** | 2.068 | -0.213 |
| **SNP + C** | 2.098 | -0.306 |
| **SNP + T** | 2.176 | -0.222 |

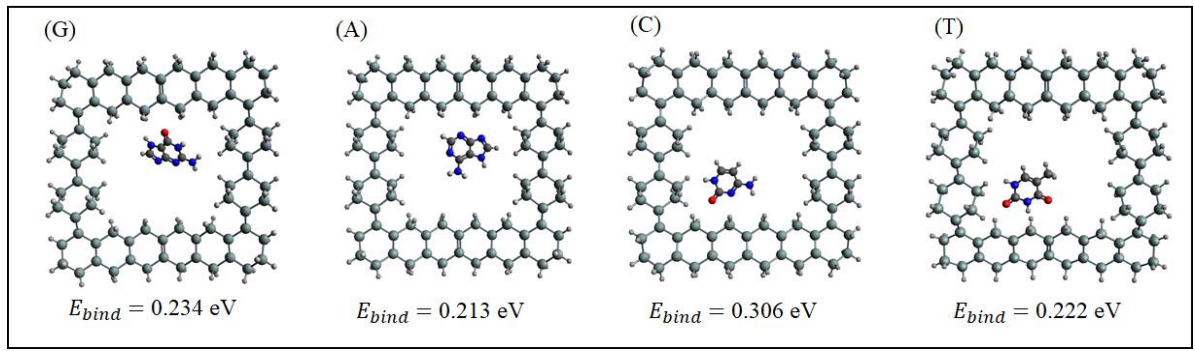

**FIG. 8.** Optimized structures of the SNP + nucleobase systems. The absolute value of the computed binding energies are shown for each base.

The optimized geometry for SNP and DNA bases is shown in **Figure 8**. In the GNP, the binding energy for each base showed C > A > G > T. For PNP, the binding energy followed the order C > G

> A > T. For SNP, the sequence is C > G > T > A. The variation of binding energy and energy gap for each base is shown in **Figure 9**.

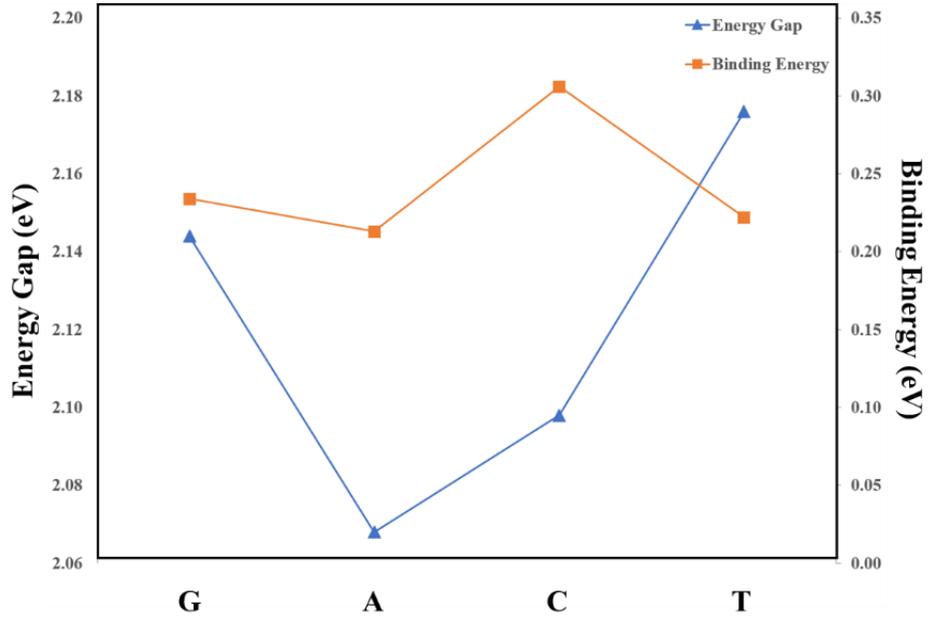

**FIG. 9**. Energy Gap (left axis) and Binding Energy (right axis) of the pristine SNP system with each DNA base placed in the pore.

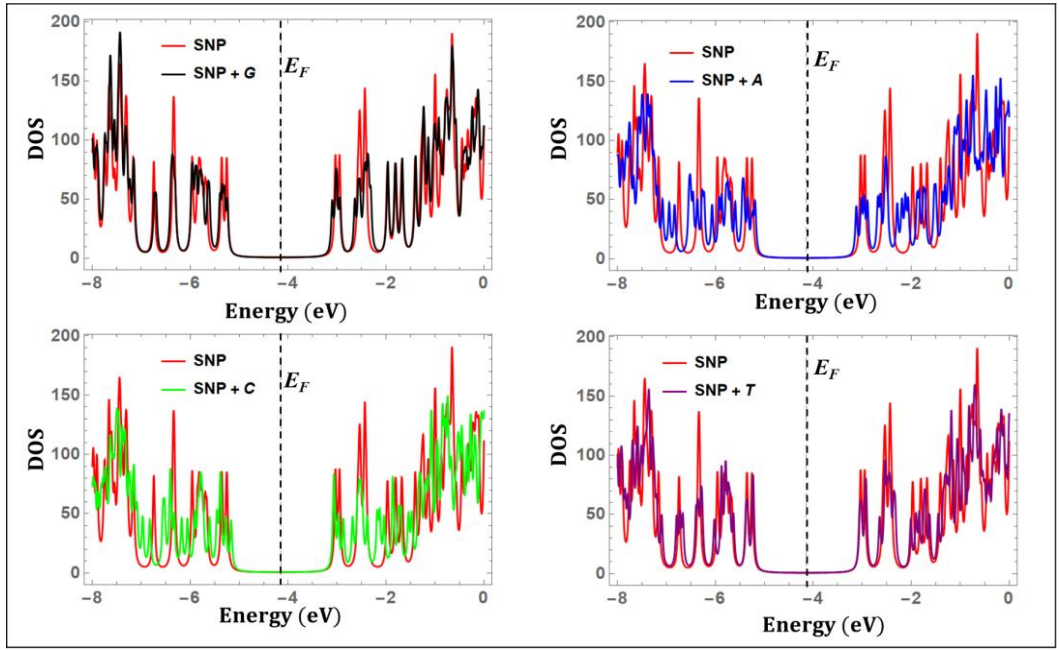

**FIG. 10.** Density of States of pristine SNP and pristine SNP + base systems. $E_F$ is the Fermi energy of pristine SNP.

From **Figure 10**, we observe that the change in band gap and DOS is unique for each base. Also, the change in band gap is large for the SNP device than the GNR device, again confirming the hypothesis that silicene might be a superior sensing material for transverse current detection compared to graphene.

**III.4 Comparing GNP, PNP, and SNP.** This research sought to explore the properties of finite-size nanoribbons from graphene, phosphorene, and silicene as DNA base detectors using the nanopore method. The pore size for GNP (1.07 nm) is comparable to the pore size for PNP (1.03 nm). The pore size for SNP was slightly larger (1.20 nm). Despite the slight differences, we can compare the general trends of the modulation in electronic properties due to interaction with different DNA bases. **Table IV** shows the binding energy and energy gap for the different materials. We observe from **Table IV** that binding energies of SNP and PNP are comparable, and lower than binding energies in GNP. To further quantify the changes in band gap, we calculated the change in band gap as the difference in band gap of the pristine system and base, and the pristine system without DNA base. **Table V** shows the change in band gap for each material. It shows that the modulations in energy gap for PNP and SNP are significant compared to GNP. Finally, **Figure 11** shows the binding energy comparison for each of the different membrane materials in the presence of each base. The lower binding energies of DNA bases in PNP and SNP suggest minimal sticking of DNA bases to pore for these materials. Hence PNP and SNP are expected to resolve the issue of pore stickiness, which is a major challenge for DNA translocation through graphene nanopore.[4,9,31]

**TABLE IV**. Energy Gap and Binding Energy for GNP, PNP, and SNP.

| Base | Energy Gap (eV) | | | Binding Energy (eV) | | |
|---|---|---|---|---|---|---|
| | GNP | PNP | SNP | GNP | PNP | SNP |
| **Pristine** | 0.221 | 3.070 | 2.219 | - | - | - |
| **G** | 0.228 | 3.083 | 2.144 | 0.888 | 0.395 | 0.234 |
| **A** | 0.230 | 2.789 | 2.068 | 0.936 | 0.307 | 0.213 |
| **C** | 0.230 | 3.046 | 2.098 | 1.063 | 0.405 | 0.306 |
| **T** | 0.231 | 3.025 | 2.176 | 0.871 | 0.207 | 0.222 |

**TABLE V.** Change in energy gap for GNP, PNP, and SNP systems due to the presence of DNA base.

| Base | $\Delta E_{gap}$ (eV) | | |
|---|---|---|---|
| | GNP | PNP | SNP |
| **G** | +0.007 | +0.013 | -0.075 |
| **A** | +0.009 | -0.281 | -0.151 |
| **C** | +0.009 | -0.024 | -0.121 |
| **T** | +0.010 | -0.045 | -0.043 |

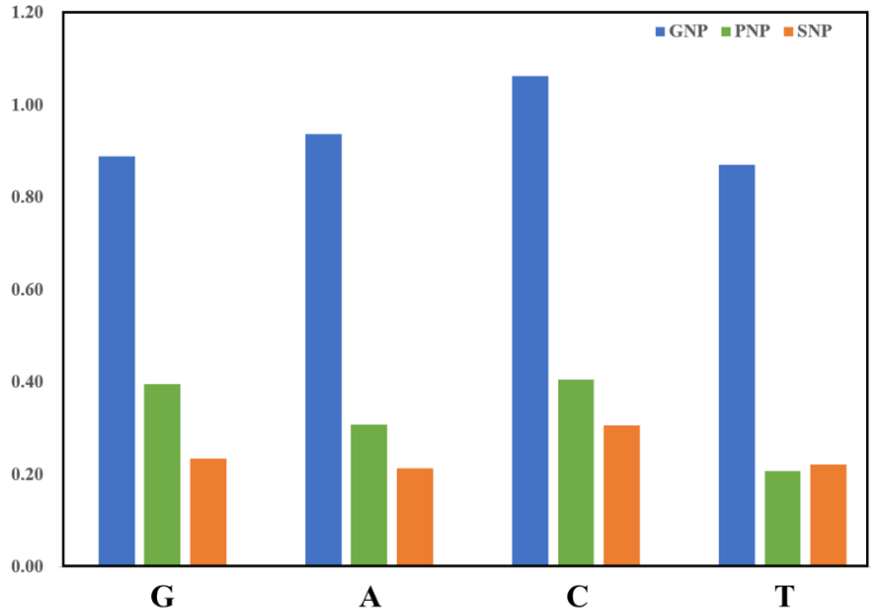

**FIG. 11**. Visualization of differences in binding energy (eV) for GNP, PNP, and SNP.

**IV. DISCUSSION**

In this study, ions and solvating water molecules were not included in our idealized simulation.[7,9,29,30] Other studies have shown that solution do play an important role. We would expect solution to change the magnitude of the binding energies and band gap, but the trend will not be changed.[39-41] For the orientation of bases, we focused only on the orientation with bases parallel to plane of nanoribbons ($\theta = 0$). Orientation does affect the binding energy and band gap of the various systems. For example, the binding energy and band gap of GNP for different orientations of DNA bases are shown in **Table VI** and **Table VII**, respectively. We observe from **Tables VI** and **VII** that the modulation in binding energy and band gap of GNP for different orientations of DNA bases is not very significant. Similar results are expected for PNP and SNP. Therefore, one would expect the trends in **Table V** and **Figure 11** to be the same for different orientations of the DNA bases.

Table VI. Binding energy of GNP for different orientations of DNA bases.

| System | $E_{bind}$ (eV) ($\theta = 0$) | $E_{bind}$ (eV) ($\theta = 45°$) | $E_{bind}$ (eV) ($\theta = 90°$) |
|---|---|---|---|
| **GNP + G** | -0.888 | -1.034 | -0.998 |
| **GNP + A** | -0.936 | -0.878 | -0.931 |
| **GNP + C** | -1.063 | -1.077 | -0.910 |
| **GNP + T** | -0.871 | -0.886 | -0.810 |

Table VII. Energy gap of GNP for different orientations of DNA bases.

| System | $E_{gap}$ (eV) ($\theta = 0$) | $E_{gap}$ (eV) ($\theta = 45°$) | $E_{gap}$ (eV) ($\theta = 90°$) |
|---|---|---|---|
| **GNP + G** | 0.228 | 0.231 | 0.227 |
| **GNP + A** | 0.230 | 0.228 | 0.228 |
| **GNP + C** | 0.230 | 0.231 | 0.234 |
| **GNP + T** | 0.231 | 0.229 | 0.229 |

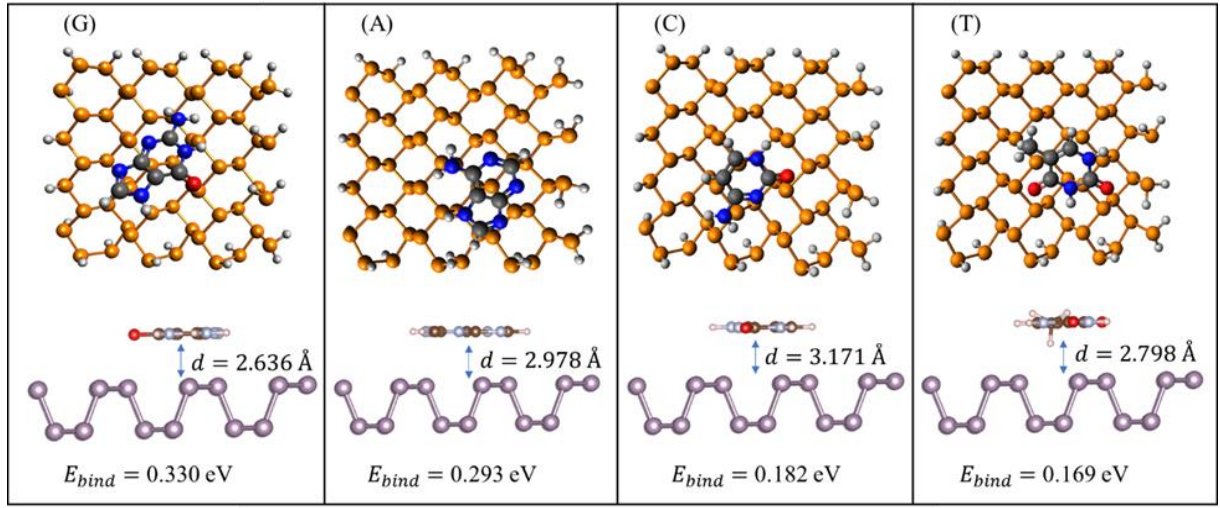

**FIG. 12**. Optimized structures of adsorption of DNA bases on the surface of phosphorene nanoribbon (PNR). The absolute value of the computed binding energies are shown for each base.

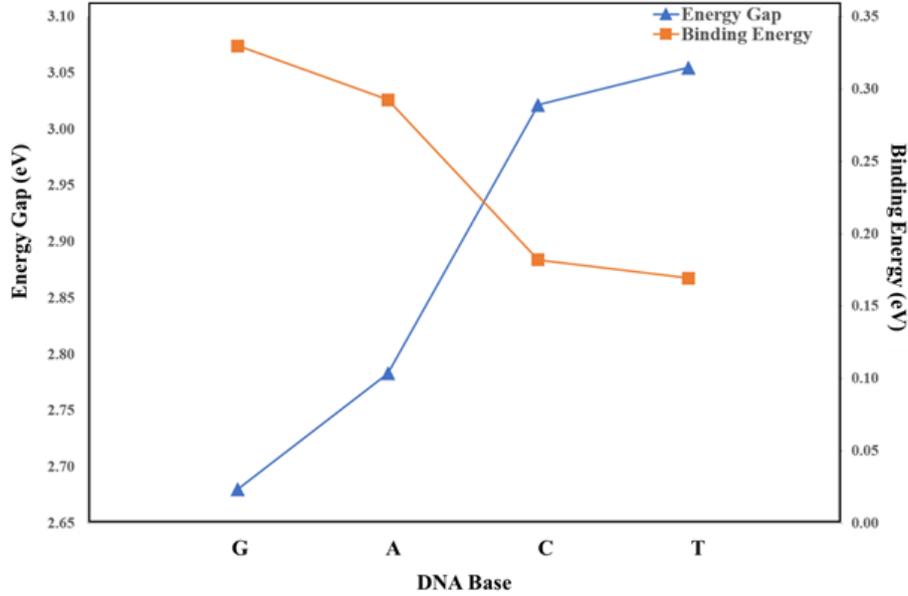

**FIG. 13.** Energy Gap (left axis) and Binding Energy (right axis) of the pristine PNR system with each DNA base placed in the pore.

Our finite-size model is a simplification of real systems. As a result, our values of binding energies obtained using Gaussian local basis set is expected to be different from bulk (infinitely long nanoribbons) values that are usually obtained with plane wave basis set. Binding energies calculated using Eq. (2) may be subjected to basis set superposition error (BSSE). For the relatively decent size of basis set and choice of model chemistry used in our calculations, the BSSE is expected to be small, and hence can only affect the magnitude of binding energies but not the general trend. To further support this argument, we report our recently obtained results for adsorption calculations for DNA bases on finite-size nanoribbons using the same basis set and level of theory that was used for the nanopore calculations (see Section II), and then compared our finite-size calculations with results that were recently published in the literature for infinitely long nanoribbons.[30] For simplicity, we present results for only the finite-size phosphorene nanoribbon (PNR). The finite-size PNR has 9 dimer lines along the zigzag direction and 3 translational periods along the armchair direction. **Figure 12** shows the optimized geometry for each base, along with corresponding binding energies (absolute values), and equilibrium adsorption heights (all bases were placed 3.0 Å above surface of nanoribbon before geometry optimization). The binding energies follow the order G > A > C > T. This same order was observed for phosphorene nanoribbons using periodic plane wave calculations.[30] The binding energies (0.169 to 0.330 eV) for the finite-size model are lower than those obtained using periodic calculations (1.75 to 2.32 eV), but the order G > A > C > T is the same for both finite-size and periodic systems.[30] The discrepancy in binding energy magnitudes is due to the difference in supercell size used for the periodic calculation. Our calculated band gap of pristine PNR is 3.038 eV. This value is higher than the value obtained using periodic calculations (~ 1.0 eV) for same width PNR.[30] This discrepancy is due to the finite-size approximation used in our model. The variation of binding energy and energy gap for each base is shown in **Figure 13**. Each base opens the band gap of the pristine PNR. Base G most significantly affects PNR and gives rise to a decrease in band gap of 0.358 eV. The change in band gap is highest in G and lowest in C and follows the order G > A > C ~ T. Interestingly, this order is also observed for interaction of DNA bases with infinitely long nanoribbons from phosphorene,[30] graphene,[42] and $MoS_2$.[9] Thus, we conclude that our choice of basis set and model chemistry appropriately captures the qualitative aspects of the systems investigated. The inclusion of BSSE correction could therefore lead to a quantitative change in binding energy for finite-size systems, but the trend is expected to remain unchanged.

## V. CONCLUSION

In summary, using DFT, we studied the interaction of DNA basis with nanopores created in finite-size nanoribbons from graphene, phosphorene, and silicene. We observe that binding energies of DNA bases using nanopores from phosphorene (0.2 – 0.4 eV) and silicene (0.2 to 0.3 eV) are similar, and generally smaller compared to graphene (0.8 – 1.1 eV). This shows that minimal sticking of DNA bases to pore is expected for phosphorene and silicene devices. The binding energy sequence was determined as C > A > G > T for graphene, C > G > A > T for phosphorene, and C > G > T > A for silicene. Furthermore, nanopores from phosphorene and silicene show a characteristic change in the density of states for each base. The band gap of phosphorene ($\Delta E_{gap} \sim 13 - 281$ meV) and silicene ($\Delta E_{gap} \sim 43 - 151$ meV) are significantly altered due to interaction with DNA bases compared to graphene ($\Delta E_{gap} \sim 7 - 10$ meV). Our findings show that phosphorene and silicene are promising alternatives to graphene for DNA base detection using advanced detection principles such as transverse tunneling current measurement.

## AUTHORS CONTRIBUTIONS

All authors contributed equally to this work.

## ACKNOWLEDGEMENT

This work was funded by the Faculty On-Campus Grant and the RCSA Grant from the Office of Research and Sponsored Programs at the University of Central Oklahoma. The author wishes to thank the College of Mathematics and Science at the University of Central Oklahoma for the CURE-STEM research funds. Computational Resources were provided by the University of Central Oklahoma Buddy Supercomputer.

## DATA AVAILABILITY

The data that support the findings of this study are available from the corresponding author upon reasonable request.

## REFERENCES

1. S. Garaj, W. Hubbard, A. Reina, J. Kong, D. Branton and J. A. Golovchenko, *Nature* **467**, 190 (2010).

2. C. A. Merchant, K. Healy, M. Wanunu, et al., *Nano Lett.* **10**, 2915 (2010).
3. K. S. Novoselov, A. Mishchenko, A. Carvalho and A. H. Castro Neto, *Science* **353**, aac9439 (2016).
4. C. Sathe, X. Zou, J. P. Leburton and K. Schulten, *ACS Nano.* **5**, 8842 (2011).
5. D. B. Wells, M. Belkin, J. Comer and A. Aksimentiev, *Nano Lett.* **12**, 4117 (2012).
6. G. F. Schneider, Q. Xu, S. Hage, et al., *Nat Commun.* **4**, 2619 (2013).
7. T. Nelson, B. Zhang and O. V. Prezhdo, *Nano Lett.* **10**, 3237 (2010).
8. W. P. Lv, M. D. Chen and R. A. Wu, *Soft Matter* **9**, 960 (2013).
9. A. B. Farimani, K. Min and N. R. Aluru, *ACS Nano.* **8**, 7914 (2014).
10. M. Graf, M. Lihter, M. Thakur, et al., *Nat Protoc.* **14**, 1130 (2019).
11. S. Z. Butler, S. M. Hollen, L. Cao, et al., *ACS Nano.* **7**, 2898 (2013).
12. B. N. Miles, A. P. Ivanov, K. A. Wilson, F. Doğan, D. Japrung D and J. B. Edel, *Chem. Soc. Rev.* **42**, 15 (2013)
13. J. Feng, K. Liu, R. D. Bulushev, S. Khlybov, D. Dumcenco, A. Kis and A. Radenovic, *Nat Nanotechnol.* **10**, 1070 (2015).
14. M. Graf, M. Lihter, D. Altus, S. Marion and A. Radenovic, *Nano Lett.* **19**, 9075 (2019).
15. K. Liu, J. Feng, A. Kis and A. Radenovic, *ACS Nano.* **8**, 2504 (2014).
16. Y. Zhao, Y. Xie, Z. Liu, X. Wang, Y. Chai and F. Yan, *Small* **10**, 4521 (2014).
17. S. Liu, B. Lu, Q. Zhao, J. Li, T. Gao, Y. Chen, Y. Zhang, Z. Liu, Z. Fan, F. Yang, et al., Adv. Mater. **25**, 4549 (2013).
18. L. Zhang and X. Wang, *Nanomaterials* **6**, 111 (2016).
19. Y. He, M. Tsutsui, S. Ryuzaki, K. Yokota, M. Taniguchi and T. Kawai, *NPG Asia Materials* **6**, e104 (2014).
20. Q. Lin, X. Zou, G. Zhou, R. Liu, J. Wu, J. Li and W. Duan, *Phys. Chem. Chem. Phys.* **13**, 12225 (2011).
21. D. Branton, D. W. Deamer, A. Marziali, et al., *Nat Biotechnol.* **26**, 1146 (2008).
22. P. Zereshki, Y. Wei, F. Ceballos, et al., *Nanoscale* **10**, 11307 (2018).
23. D. Jose and A. Datta, *Acc. Chem. Res.* **47**, 593 (2014).
24. Q. Wei and X. Peng, *Appl. Phys. Lett.* **104**, 251915 (2014).
25. X. Y. Zhou, R. Zhang, J. P. Sun, Y. L. Zou, D. Zhang, W. K. Lou, F. Cheng, G. H. Zhou, F. Zhai, and K. Chang, *Scientific Reports* **5**, 12295 (2015).
26. Tony Low, A. S. Rodin, A. Carvalho, Yongjin Jiang, Han Wang, Fengnian Xia, and A. H. Castro Neto, *Phys. Rev. B* **90**, 075434 (2014).
27. Y. Chen, R. Ren, H. Pu, J. Chang, S. Mao, and J. Chen, *Biosensors & Bioelectronics*, **89**, 505 (2017).
28. D. Cortés-Arriagada, *J. Phys. Chem. C* **122**, 4870 (2018).
29. R. L. Kumawat, P. Garg, S. Kumar, and B. Pathak, *ACS Appl. Mater. Interfaces* **11**, 219 (2019).
30. R. L. Kumawat and B. Pathak, *J. Phys. Chem. C* **123**, 22377 (2019).
31. S. J. Heerema and C. Dekker, *Nat Nanotechnol.* **11**, 127 (2016).
32. Gaussian 16, Revision C.01, Frisch, M. J. et al., Gaussian, Inc., Wallingford CT, 2016.
33. This project made use of the Buddy Supercomputer at the University of Central Oklahoma, which was funded by National Science Foundation award number ACI-1429702.
34. J. Prasongkit, A. Grigoriev, B. Pathak, R. Ahuja and R. H. Scheicher, *Nano Lett.* **11**, 1941 (2011).
35. B. O. Tayo and S. V. Rotkin, *Phys. Rev. B* **86**, 125431 (2012).
36. D. Jose and A. Datta, *Acc. Chem. Res.* **47**, 593 (2014).
37. Z. Ni *et al.*, *Nanoscale* **6**, 7609 (2014).
38. S. Cahangirov, M. Topsakal, E. Aktürk, H. Šahin, and S. Ciraci, *Phys. Rev. Lett.*, **102**, 236804 (2009).
39. G. T. Feliciano, C. Sanz-Navarro, M. D. Coutinho-Neto, P. Ordejon, R. H. Scheicher, R. H. and A. R. Rocha, *J. Phys. Chem. B* **122**, 485 (2018).
40. J. Lagerqvist, M. Zwolak, and M. Di Ventra, *Nano Lett.* **4**, 779 (2006).
41. J. Lagerqvist, M. Zwolak, and M. Di Ventra, *Biophys. J.* **93**, 2384 (2007).
42. J. –H. Lee, Y. –K. Choi, H. –J. Kim, R. H. Scheider, and J. H. Cho, *J. Phys. Chem. C* **117**, 13435 (2013).